\documentclass[%
 reprint,
superscriptaddress,
amsmath,amssymb,
 aps,
 prl,
floatfix,
]{revtex4-2}

\usepackage{physics}
\usepackage{color}
\usepackage{graphicx}
\usepackage{dcolumn}
\usepackage{bm}
\usepackage{calc}
\usepackage{verbatim}
\usepackage{float}

\usepackage[mode=buildnew]{standalone}
\begin{document}

\title{Spontaneous symmetry breaking in nonlinear superradiance}

\author{Nikolai D. Klimkin}
\email{Nikolay.Klimkin@mbi-berlin.de}
 \affiliation{Max Born Institute, Max-Born Stra{\ss}e 2A, 12489 Berlin, Germany}
\author{Misha Ivanov}%
\affiliation{Max Born Institute, Max-Born Stra{\ss}e 2A, 12489 Berlin, Germany}
\affiliation{Institute of Physics, Humboldt University Berlin, 12489 Berlin, Germany}
\affiliation{Technion – Israel Institute of Technology, 3200003 Haifa, Israel}

\begin{abstract}
    Creation and manipulation of non-classical states of light is rapidly becoming the focus of modern attosecond science. Here, we demonstrate numerically how interaction with such states can trigger the emergence of a many-body system with spontaneously broken symmetry 
    by considering a modification of the well-known problem of superradiance encountered already by Dicke. Similarly to him, we investigate photon emission by ensembles of indistinguishable atoms. In contrast to him, however, we leverage symmetry-based selection rules to suppress emission of single photons by single atoms. A steady state is therefore only reached following a spontaneous transition into a collective symmetry-broken state of atoms and photonic modes. This transition permanently locks the atomic dipoles to the quantum field experienced by the system at a particular instant, transforming the entire setup into a potent quantum sensor reproducing the phase of the recorded quantum fluctuation.
\end{abstract}

\maketitle

Superradiance~\cite{dicke1954coherence}  is the textbook example of a cooperative emission phenomenon. While recent work~\cite{tziperman2023quantum,pizzi2023light} suggests that superradiance can acquire quantum properties, for the most part it has been regarded~\cite{gross1982superradiance} as classical radiation. Here we show how atoms driven by an intense classical light while confined in a strongly detuned cavity can spontaneously develop many-body entanglement with the cavity modes and each other, generating light at frequencies not otherwise present in their emission spectrum, with the properties of this light reproducing the properties of the cavity's quantum field. 

Experiments in strong-field physics are increasingly focusing on the quantum optical properties of nonlinear optical emission produced by strongly-driven atoms and solids~\cite{lewenstein2021generation, theidel2024evidence}. The required theoretical treatment faces many challenges. Already the sheer range of new frequencies and the very large number of  photons generated by strongly driven matter exhaust the capabilities of common computational techniques. Likewise, polaritonic chemistry deals with entangled states of a few photons and molecules in cavities (e.g. ~\cite{ribeiro2018polariton, silva2020polaritonic, sandik2025cavity}) and struggles to adapt to arbitrarily high photon counts. Our work resolves this problem and shows how scalable quantum states of light can be engineered by tailoring the parameters of atoms and the cavity, opening a new field of cooperative nonlinear superradiance.

In conventional superradiance, collective emission occurs at frequencies close to those of individual atoms~\cite{schmidt2023collective, tebbenjohanns2024predicting, yatsenko2025photon}. This is also true in recent experiments on collective emission in trapped ions~\cite{richter2023collective, verde2025spin}. Many-body dynamics in superradiant setups has also been investigated in a recent experiment~\cite{kersten2026self}, where they have been shown to give rise to continuous emission over macroscopic time scales. 

Here, however, we analyze a dramatically different case, when the cavity modes are detuned far away from the atomic resonance. Emission into such modes requires a nonlinear cooperative transition involving several atoms toggling their states in a correlated fashion. That is, for atoms with a resonant frequency $\omega_0$ and a photonic mode with frequency $\omega = (m/n) \omega_0  $ for integer $m \neq n$, $m$ atoms must transition from the excited to the ground state to emit $n$ disentangled photons. 

Demonstrated in experiment to be a promising single-photon source~\cite{muller2015coherent}, emission into detuned photonic modes poses formidable theoretical challenges. 
The conventional approaches, which view the entire mode continuum through an effective perturbative description \cite{pizzi2023light, yi2025generation}, or use a non-perturbative description but constrain the entire spectrum to a few effective modes \cite{tziperman2023quantum}, are insufficient when considering large arrays of strongly driven emitters strongly coupled to photonic modes. Here, one faces the challenges of (i) describing strongly correlated many-body dynamics of quantum emitters developing through the coupling to a common light mode~\cite{muller2025genuine}, (ii) accounting for possible symmetry breaking by strong quantum fluctuations in emitters~\cite{stitely2023quantum}, all in conjunction with (iii) arbitrarily large quantum states of light generated
via nonlinear optical response to a strong external driving field. The broad emission spectra and the large numbers of the emitted photons typically force one to use a Markov-regime description of the emitters' interaction with the quantum vacuum. The Markov regime would rule out any emitter-mode entanglement, and by extension generation of nontrivial states of light. Our approach (see Supplementary information for details) overcomes these challenges.

\begin{figure}[b]
    \centering
    \includegraphics[width=\linewidth]{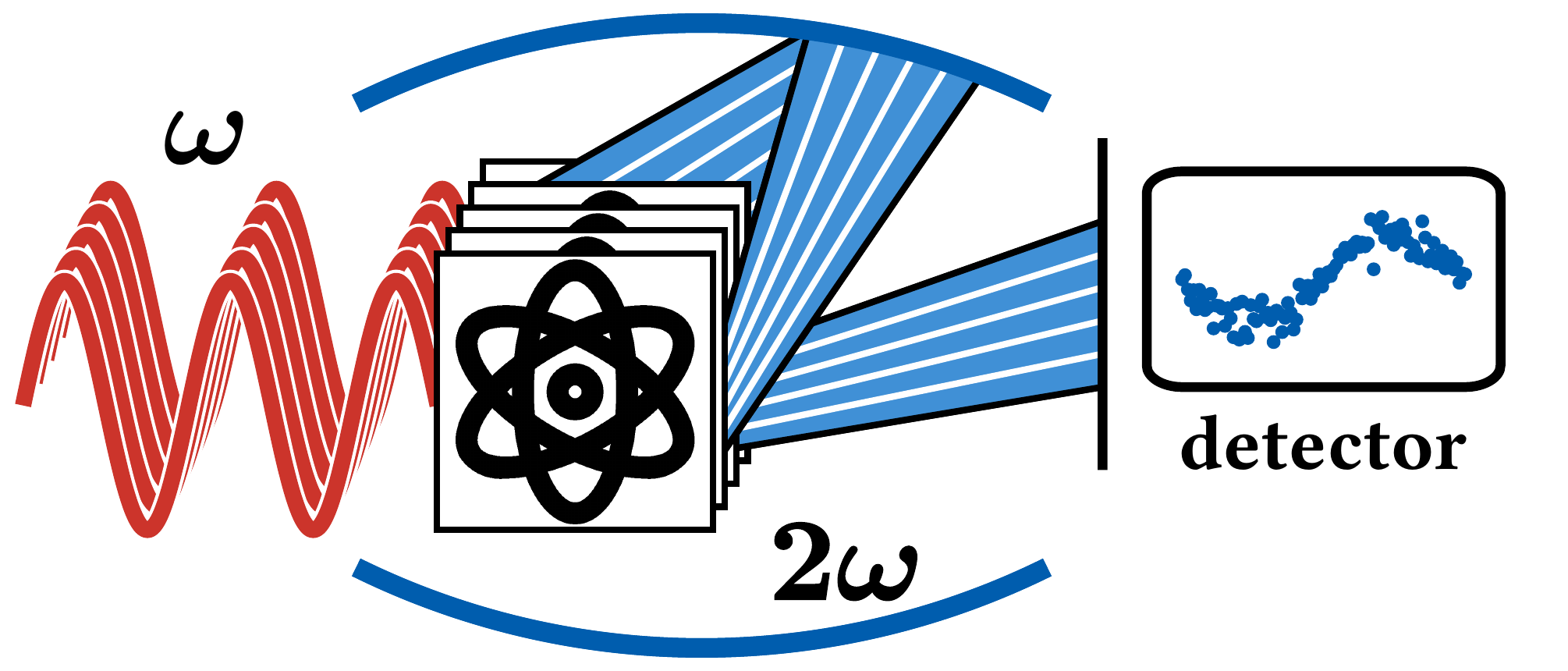}
    \caption{(color online) Basic design of our setup. The classical radiation (red), resonant in our case with the atomic transitions, drives ultrafast currents in all atoms independently. This causes them to emit frequency-converted photons (blue) entangled with the emitters. The electromagnetic radiation confinement exhibited by e.g. a cavity (blue lines) causes these photons (blue beams with white lines) to linger, allowing them to either be absorbed by the same or another atom, or leave the confining medium. In the latter case, they can be observed by a classical detector positioned outside (right). By detuning the cavity's resonant frequency far from the atomic resonance, we restrict the emission of observable light to groups of atoms.}
    \label{fig:fig1}
\end{figure}

\begin{figure}[b]
    \centering
    \includegraphics[width=\linewidth]{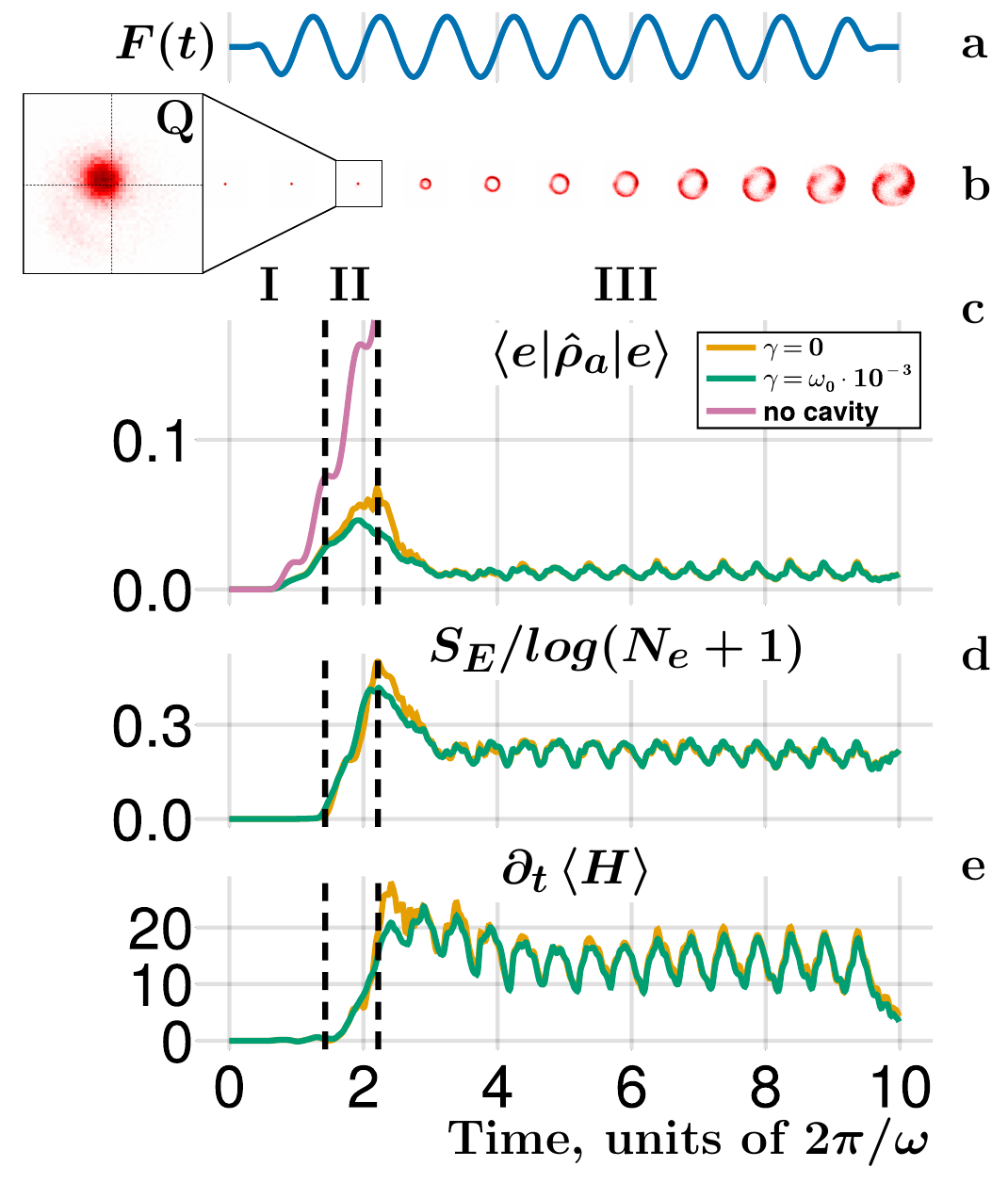}
    \caption{(color online) Collective emission dynamics. All horizontal axes correspond to the time measured in laser cycles $T_0 = 2\pi/\omega$. The yellow line corresponds to the evolution of emitters in the presence of the cavity with no additional side channel decoherence. The green line shows evolution which includes an additional decoherence channel. The purple line in (b) shows evolution with the decay channel, but without the cavity. \textbf{(a)} Husimi Q function for discretized mode 7 plotted for different moments in time. Indistinguishable from vacuum at first, it spreads out into a manifestly non-Gaussian shape at onset of phase III. \textbf{(b)} The waveform of the driving field $F(t)$, set in resonance with a single atomic transition. Inset shows the Husimi function at $t=2$ laser cycles. The dashed crosshairs show the origin $\alpha = 0$, demonstrating the displacement of the light state and the emerging density halo; \textbf{(c)} Average fraction of excited atoms $\expval{\hat{\rho}_a}{e}$;
    \textbf{(d)} Von Neumann entropy (VNE) of the emitter state, normalised to the VNE of the respective maximally mixed state $\log{\left(N_e+1\right)}$;
    \textbf{(e)} Average photonic energy production rate with the energy given by the Hamiltonian (\ref{hb-k}); 
    }
    \label{fig:multiplot-2h}
\end{figure}

Fig.~\ref{fig:fig1} depicts our system of many $N_e\gg1$ atoms (here, $N_e=40$) inside a cavity driven by a strong classical field. By having a finite bandwidth, this cavity incorporates a unitary way for the photons to separate themselves from the emitters. A similar effect can be achieved in e.g. a waveguide or a photonic crystal with a sufficiently low group velocity~\cite{busch1999liquid, busch2000radiating}. The atoms are modeled as two-level systems with a resonance frequency $\omega_0$, subjected to a classical driving force. The driving field shown in Fig.~\ref{fig:multiplot-2h}(a) is resonant with the atomic transition: $\omega = \omega_0$. The cavity, however, is strongly off-resonant: 
its resonance is centered at $\epsilon = 2\omega_0$. As second harmonic generation
is forbidden in a centrally symmetric medium, the cavity suppresses uncorrelated emission by individual atoms.

The cavity with the Hamiltonian 
\begin{equation}\label{hb-k}
   H_B = \int d \omega \omega \ \hat{a}_\omega^\dagger \hat{a}_\omega 
\end{equation}
is modeled as a waveguide, with the frequency modes described using a standard tight-binding approximation with the next-neighbor coupling constant $h$. Within this approximation, the cavity eigenstates are excitations indexed by a wavenumber $k \in [0, \pi)$:
\begin{equation}\label{omega-k}
 \omega(k) = \epsilon + 2h \cos{k} ; \ \
 c(k) = h \sqrt{2 \pi} \sin{k} \ \ .
 \end{equation}
Here the coefficient $c(k)$  describes the coupling of the atoms to the cavity,
\begin{equation}\label{hi-k}
 \hat{H}_I = h \hat{S}_+ \int d\omega c(\omega) \hat{a}_\omega + \text{h.c.} 
 \end{equation}
with the coupling constant $h$. The operator $\hat{S}_+$ creates an excitation in one of the atoms. 
In the calculations, we set $h=0.2\omega_0$.
For more details on the numerical implementation, see the Supplementary.

A head-on approach to this problem is ill-advised. An exact diagonalization solution of the time-dependent Schrödinger equation would be prohibitively expensive due to the fact a continuously driven system emits photons continuously, which in its turn causes the required memory space to also scale up continuously. Given the broadband setup we have chosen, storing a quantum state beyond 5-10 photons becomes infeasible. For this reason, we use instead the Hierarchy of Pure States (HOPS) method~\cite{hartmann2017exact, polyakov2019dressed}. Unlike exact diagonalization, HOPS treats the joint state of the emitters and the light field as a sequence indexed by a number $\xi$, consisting of classical amplitudes $\bm{\alpha}_\xi$ and their respective emitter state projectors $\ketbra{\psi(\bm{\alpha}_\xi)}$. The $\bm{\alpha}_\xi$ have the physical meaning of coherent states measured by an external detector. Mathematically, they are draws from the light field's multimode Husimi function $Q(\bm{\alpha})$. By being c-numbers, $\bm{\alpha}$, and the respective photon counts, can be scaled arbitrarily high without incurring any costs on memory space. 

In addition to the so-called ''real'' photons that may be detected, HOPS utilizes ''virtual'' photons that may not. But the virtual photons influence the atoms' dynamics nonetheless. In Fig.~\ref{fig:switch} and the attached discussion, we consider briefly the role played by the virtual photons. In our simulations, we may vary the number of the virtual photons we account for. As we increase this number, HOPS converges to the exact solution. 

The choice to center the cavity spectrum at the second harmonic of the driver is due to the following considerations. 
On the one hand,  individual atoms driven by a resonant classical field and preserving the inversion and time shift symmetry cannot emit the second harmonic. Indeed, the uncorrelated second harmonic emission by individual atoms remains negligible in our case.

On the other hand, many systems with a large amount of degrees of freedom tend toward an equilibrium. In our case, however, the atoms can only reach such equilibrium with the electromagnetic field when their gain arising from photon absorption from the driving field evens out with the loss happening due to the emission. Given that the individual atom's second harmonic emission amplitude is negligible, reaching an equilibrium becomes impossible for uncorrelated atoms. This way, we force the system to undergo a collective spontaneous symmetry-breaking transition into a correlated state where the second harmonic can be emitted efficiently.

The results demonstrating this behavior for an
ensemble of $N_e=40$ atoms driven by the waveform in Fig.~\ref{fig:multiplot-2h}(a) and coupled to the cavity are shown in Fig.~\ref{fig:multiplot-2h}
(b-e). Panel (b) shows a typical Husimi function for one of the modes within the cavity spectrum in the vicinity of the second harmonic (see Supplementary information for similar distributions for other cavity modes). Indistinguishable from the vacuum state at first, represented by the dot in the middle, the mode spreads out into a 
manifestly non-Gaussian, non-classical shape. Thus, the system of emitters, which is initially unable to emit the second harmonic, finds a way to generate it. 

The underlying dynamics is investigated in panels (c-e). 
The evolution in Fig.~\ref{fig:multiplot-2h} is divided into three phases, marking the distinct interaction regimes. We discuss them  after describing the quantities plotted in Fig.~\ref{fig:multiplot-2h}.

Yellow lines in panels (c-e) show the dynamics of the atoms coupled only to the cavity modes, while the green curves show the same in the presence of an additional Markovian decoherence into a side channel with the rate $\gamma = \omega_0 \cdot 10^{-3}$. The dashed line in panel (c) shows the dynamics of an atom without the cavity, which undergoes  the typical Rabi oscillation. 

Panel (d) shows the normalized Von Neumann entropy (VNE) $S_E = -\text{tr}\left(\hat{\rho}_a \log \hat{\rho}_a\right)$ of the emitters' partial density matrix $\hat{\rho}_a$, obtained as described in the Supplementary. For evolution with no Markovian decoherence channel, the joint emitter-photon state $\ket\Psi$ remains pure, with zero $S_E$. Thus, Fig.~\ref{fig:multiplot-2h}(d) shows strong entanglement between the photons and the emitters. Similar behavior is found in the presence of the Markovian decay channel.

To compare the system-side with the photon-side dynamics, in Fig.~\ref{fig:multiplot-2h}(e) we plot the photonic energy production rate, i.e. the average value of $\hat{H}_B$ Eq.(\ref{hb-k}).

We now turn to the dynamics in panels (b-e). In phase I, in contrast to the unimpeded Rabi oscillations (dashed line in panel (c)),  the actual excitation dynamics in Fig.~\ref{fig:multiplot-2h}(c) shows abortive Rabi oscillations, mirrored  by suppressed emitter-photon entanglement in panel (d). The amount of real photons being emitted in phase I remains negligibly small, see Fig.~\ref{fig:multiplot-2h}(b, e). The initial emission is thus characterized by a gridlock. While interaction with the cavity is significant in phase I, causing the excitation profile (c) to deviate dramatically from its Rabi counterpart, it is only mediated by virtual photons, localized around the atoms and unable to reach a distant detector. Once the correlations are sufficiently developed in phase II, the emission gridlock is broken in phase III.

At the onset of phase II,  the correlations shown in Fig.~\ref{fig:multiplot-2h}(d) begin to surge. Simultaneously, the energy production rate in Fig.~\ref{fig:multiplot-2h}(e) rises above zero, indicating the emission becoming unlocked. 

Following a rise in entanglement characterizing phase II, the emitter and the photons equilibrate, continuing to evolve in a less entangled manner in phase III, delimited by the peak VNE value. Naïvely, one may assume this would cause the state of emitted light to be closer to a coherent state, as is the case when the emitter-photon entanglement is negligibly small at all times. However, as seen in Figure~\ref{fig:multiplot-2h}(b), this is manifestly not the case. The mechanism behind this distinction is important enough to give this work its name, the spontaneous symmetry breaking. Namely, our numerical simulations show that the transient entanglement surge in Figure~\ref{fig:multiplot-2h}(c) alters permanently every trajectory of both the emitters and the photonic field. Small vacuum fluctuations registered at the onset of phase III are imprinted on the emitter dynamics, and reproduced by their subsequent emissions. An entire avalanche of emission is triggered by an initial prod from the weak vacuum fluctuations. Harmonic emission is therefore gated -- not by the classical pulse, but by quantum vacuum.

\begin{figure}[b]
    \centering
    \includegraphics[width=\linewidth]{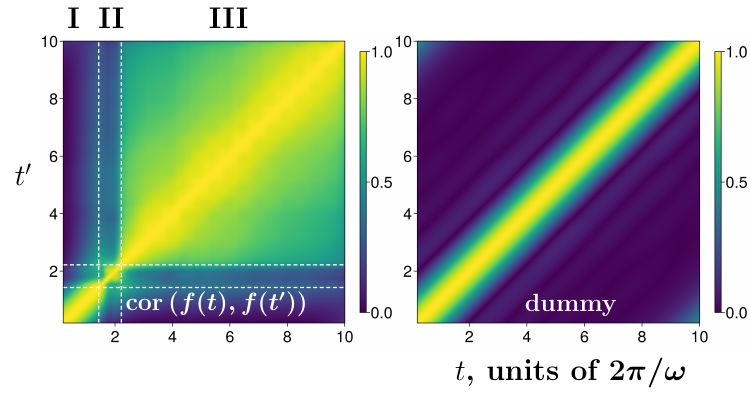}
    \caption{(color online) \textbf{(left)} Correlation between the real photonic fields $f$ for different times $t$, $t'$. The time $t$ corresponds to the horizontal, and $t'$ to the vertical axis. \textbf{(right)} Correlation absent light-matter interaction (dummy plot).}
    \label{fig:corr}
\end{figure}

\begin{figure}[b]
    \centering
    \includegraphics[width=\linewidth]{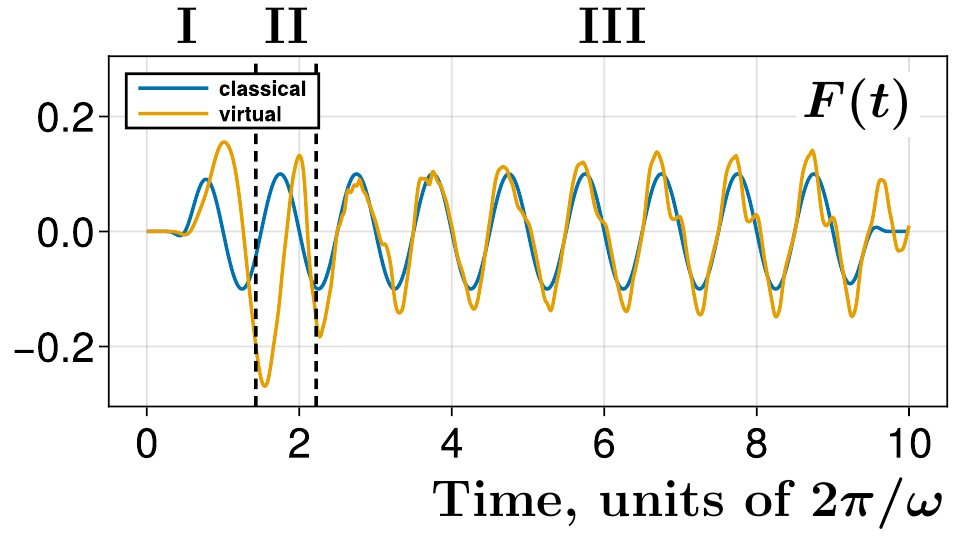}
    \caption{(color online) Virtual photonic field $\expval{\hat{F}(t)}$ (yellow) plotted alongside the classical external field (blue) with its sign inverted $-F(t)$. 
    }
    \label{fig:switch}
\end{figure}

To support this claim, Figure~\ref{fig:corr} shows the absolute pairwise Pearson correlations between stochastic quantum fields $f (t)$ calculated at different times $t, t'$ (see the Supplementary.) We are tracking how fast the correlations fade between the instantaneous field values. We plot side-by-side the actually observed temporal correlations for a cavity field interacting with the emitters, and a dummy plot depicting these correlations for a non-interacting cavity field. We observe in the dummy plot that on their own, these correlations are short-lived. Likewise, at early times, the correlation time is also short for the interacting system. However, at the onset of phase III $t_3$, i.e. the highest emitter-photon entanglement, we observe that all subsequent field values $f(t'>t_3)$ become correlated over extremely long times to $f(t_3)$. The atomic emissions synchronize with the quantum vacuum fluctuations, the resulting correlation time rising drastically. Now in lockstep with the quantum field, our emitters keep reproducing the field they experienced at the transition, giving rise to a typical spontaneous symmetry breaking setup.

In phase III, the radiation is continuously emitted in phase with the vacuum fluctuations experienced by the emitters at onset. It equilibrates gradually with the emitters' excitation by the laser driver. The properties of this outgoing radiation, which in our simulations contain many hundreds of photons,  becomes non-classical, as suggested by Fig.~\ref{fig:multiplot-2h}(b). The intensity-intensity correlation functions $g^{(n)}(0)$ (Fig.~\ref{fig:gn-sup} in the Supplementary) confirm this assessment. They fluctuate near unity in spite of the manifestly non-Gaussian Husimi distributions, and reach slightly into the manifestly non-classical region of $g^{(n)}(0)<1$ in spite of the strong emitter-mode entanglement persisting in phase III. The outgoing radiation is therefore significant because on the one hand, it's triggered by the quantum field registered by the emitters at one point, and remains correlated with it for many subsequent laser cycles. But at the same time, this radiation does not rely on an existing inverted state of the emitters~\cite{kersten2026self}, instead converting  incoming classical radiation into quantum harmonic emission. The combination of both provides an avenue for a powerful quantum sensing protocol. The fact the outgoing radiation exhibits mild non-classical properties serves as a proof of concept rather than a proposal for Fock state generation.

Finally, Fig.~\ref{fig:switch} provides an additional perspective of the physical meaning behind the three interaction regimes involved. There, we plot the expected virtual field $F_\text{virt}$, calculated as explained in the Supplementary, alongside the classical laser field. Following the initial transition, the new emitter-photon equilibrium is defined, as the plots in Fig.~\ref{fig:multiplot-2h} imply, by the absence of excitation. As Fig.~\ref{fig:switch} reveals, this corresponds to the external resonant field being canceled out by the virtual feedback field induced by the emission dynamics. 

Our method is easily generalizable to arbitrary mode setups. It is built around the distinction between "real" and "virtual" excitations, allowing for a natural separation of photons which can be registered in the far field, and ones which cannot, e.g. due to being trapped in localized bound states. Thanks to our broadband model describing the cavity, adding irreversible, i.e. Markovian loss is not required. The atoms can  exchange photons, entangle themselves, align their dynamics with the quantum field, and then transfer the photons outside the cavity, before irreversible loss of information has occurred.

One platform warrants our particular attention as a potential proving ground.
An ensemble of alkaline atoms trapped within an optical cavity such that their regular emission is suppressed by the Purcell effect can be excited with a moderately intense ($10^{11} - 10^{12} \text{ W/cm}^2$) laser source. These can be expected to undergo a Rabi oscillation, begin exchanging virtual photons, transition to a collective entangled state, and emit radiation bursts in a cooperative way, as predicted by this work. The relatively weak coupling to cavity modes can be offset by the potentially very large number of atoms involved, even at modest gas pressures.

Thanks to the new addition to the toolset applied to the problem of quantum harmonics generation, our work shows how non-perturbative interactions can reveal the fluctuations of a quantum state of light with an arbitrary magnification. However, not only the measurement, but also the manipulation of quantum states of light is held back by the insufficiency of computational methods used to describe their non-perturbative interaction with matter. We believe our work bridges this gap.
\smallskip

\begin{acknowledgments}
N.K. acknowledges support of CRC 1477 "Light-Matter Interactions at Interfaces", Project No. 441234705. M.I. acknowledges Project No. 545591821, IV 152/11-1.
N.K. acknowledges Evgeny A. Polyakov's clarifications of Ref.~\cite{polyakov2019dressed}, Vladislav Sukharnikov and Stasis Chuchurka's advice on the HOPS method, and useful advice from Johannes Feist. M.I. and N.K. acknowledge many enlightening discussions with Alexey Gorlach, Ido Kaminer, Moti Segev, and Stefanos Carlström. In addition, N.K. acknowledges Moti Segev's help in maintaining and continuing this work by assisting his reaching shelter during an inbound missile alarm. 
\end{acknowledgments}

\nocite{klimkin_2026_20056341}
\nocite{klimkin2025codebase}

\bibliography{refs}

\appendix
\section{\label{app:numset}Numerical setup}
The non-Markovian method we adopt is a modification of the Hierarchy of Pure States (HOPS) \cite{hartmann2017exact}. In addition to giving us access to field observables, this method, as argued in \cite{polyakov2019dressed}, is exceptionally well-suited for dynamically-driven systems. We implement the HOPS method as described in \cite{polyakov2019dressed}. Our implementation allows us to investigate the mutual interaction of an arbitrarily large number of emitters with a narrowband electromagnetic vacuum. We obtain the Husimi function of the generated light and then quantify its properties by computing its antinormally-ordered quantum correlators.

Our joint system includes a small reduced system (S), a large bath of non-interacting electromagnetic modes (B), and an interaction term connecting them:
\begin{equation}\label{eq:hgen}
 \hat{H} = \hat{H}_S + \hat{H}_I + \hat{H}_B 
 \end{equation}
We model the photonic bath (B) using a tight-binding-like Hamiltonian for a semi-infinite chain of bosonic states localised on evenly spaced sites numbered by $j$, connected to one another by a nearest-neighbor hopping. The Hamiltonian is:
\begin{equation}\label{eq:tbbh}
 \hat{H}_B = \epsilon \sum_{j=1}^\infty \hat{a}^\dagger_j \hat{a}_j + h \sum_{j=1}^\infty \hat{a}^\dagger_j \hat{a}_{j+1} + \text{h.c.} 
 \end{equation}
This photonic chain interacts with a reduced system (S) made up by an array of $N_e$ atoms with $n$ orbitals. As elaborated in \cite{pizzi2023light}, its Hamiltonian can be written in terms of effective bosonic creation and annihilation operators $\hat{A}^\dagger_m$, $\hat{A}_m$, with $\left[\hat{A}_m, \hat{A}^\dagger_{m'}\right] = \delta_{m m'}$, as:
\begin{equation}\label{ham-s}
 \hat{H}_S = \sum_{m=1}^n \epsilon_m \hat{A}^\dagger_m \hat{A}_m + F(t) \sum_{m=1}^n \sum_{m'=1}^n d_{m m'} \hat{A}^\dagger_m \hat{A}_{m'} 
 \end{equation}
The Hilbert space for this Hamiltonian corresponds to $n$ effective bosonic modes filled by a total of $N_e$ atoms. 

If each emitter has $n=2$ and its  basis is chosen such that $\epsilon = (\omega_0/2, -\omega_0/2)$, $\hat{d} = d_0 \hat{\sigma}_x$, the total Hamiltonian can be written in a familiar form:
\begin{equation}\label{hs-S}
 \hat{H}_S = (\omega_0/2) \hat{S}_z + d_0 F(t) \hat{S}_x 
 \end{equation}
where $\hat{S}_x$, $\hat{S}_z$ belong to a set of operators for describing the collective pseudospin of the emitters:

\begin{eqnarray}
    \hat{S}_+ &:=& \hat{A}_2^\dagger \hat{A}_1\nonumber\\
    \hat{S}_- &:=& \hat{A}_1^\dagger \hat{A}_2\nonumber\\
    \hat{S}_x &:=& \hat{S}_+ + \hat{S}_- \\
    \hat{S}_y &:=& i (\hat{S}_+ - \hat{S}_-) \nonumber\\ 
    \hat{S}_z &:=& \hat{A}_1^\dagger \hat{A}_1 -\hat{A}_2^\dagger \hat{A}_2\nonumber
\end{eqnarray}

The interaction (I) operator is 
\begin{equation}
 \hat{H}_I = h \hat{S}_+ \hat{a}_1 + \text{h.c.} 
 \end{equation}
 
A generic photonic bath Hamiltonian is given in terms of the continuous-frequency creation and annihilation operators $\hat{a}^\dagger(\omega)$, $\hat{a}(\omega)$ following the conventional commutation relations:

\begin{equation}
    \left[\hat{a}(\omega), \hat{a}^\dagger(\omega')\right] = \delta(\omega - \omega')
\end{equation}

\begin{equation}\label{hb-omega}
 H_B = \int dk \omega \hat{a}^\dagger (\omega) \hat{a}(\omega)
\end{equation}

The Hamiltonian (\ref{hb-omega}) does not include a density of states, and does not account for the bath being spectrally limited. Instead, both the density of photonic states and the frequency-dependent photon mode coupling are described by the generic frequency-dependent coupling coefficient $c(\omega)$ in the integral below:

\begin{equation}\label{hi-omega}
 \hat{H}_I = h \hat{S}_+ \int d\omega c(\omega) \hat{a}(\omega) + \text{h.c.} 
 \end{equation}

For a photonic energy band modeled in the tight-binding approximation as given by (\ref{eq:tbbh}), we Fourier transform along the semi-infinite $m$ dimension to arrive at non-interacting delocalized modes indexed by the wavenumber $k \in [0, \pi)$. Instead of $c(\omega)$, we then work with a $c(k)$ and $\omega(k)$ defined in terms of $k$ as a parameter. For $\epsilon$, $h$ matching the ones used in (\ref{eq:tbbh}), their exact form is:
 
\begin{equation}
    \omega(k) = \epsilon + 2h \cos{k} 
\end{equation}
\begin{equation}
    c(k) = h \sqrt{2 \pi} \sin{k} 
\end{equation}

To be useful for a practical simulation, the mode continuum needs to be discretized into $N_m$ modes, numbered by an index $\nu = 1\ldots N_m$. We choose the modes such that the grid over $k$ is uniform:

\begin{equation}
    k_\nu = \frac{\pi \nu}{N_m + 1}
\end{equation}

We then set $\omega_\nu = \omega(k_\nu), c_\nu = c(k_\nu)$. The Hilbert space for quantum photons is then truncated at $N_p$ total photons -- i.e. we include every quantum photon state $\ket{n_1, \ldots, n_{N_m}}$ for $\sum_{\nu=1}^{N_m} n_\nu \leq N_p$. Then, labeling the cavity-side operator in (\ref{hi-omega}) as $\hat{b}$:

\begin{equation}
    \hat{b} := \int d\omega c(\omega) \hat{a}(\omega)
\end{equation}

In order to preserve commutation relations between the discretized creation and annihilation operators $\left[\hat{a}_\nu, \hat{a}^\dagger_{\nu'}\right] = \delta_{\nu\nu'}$, we discretize the operator $\hat{b}$ as:

\begin{equation}\label{eq:bd}
    \hat{b} = \sum_{\nu=1}^{N_m} c_\nu \hat{a}_\nu \sqrt{\Delta \omega_\nu}
\end{equation}

Respectively, the discretized field Hamiltonian is:

\begin{equation}
    \hat{H}_B = \sum_{\nu=1}^{N_m} \omega_\nu \hat{a}_\nu^\dagger \hat{a}_\nu
\end{equation}

\section{\label{app:calcobs}Calculating observables}

By being stochastic, our method yields the overall system-bath state in the form of many statistical samples numbered by an index $\xi$ running from $1$ to the total number of samples $N_\text{batch}$. Each of these samples contains a wavefunction $\ket{\Psi_\xi}$ and a draw from the bath's Husimi distribution $Q(\bm\alpha)$, designated as $\bm\alpha_\xi$, correlated with $\ket{\Psi_\xi}$. The wavefunction $\ket{\Psi_\xi}$ will then be called conditional, or equivalently, conditioned on $\bm{\alpha}_\xi$. 

The Hilbert space of the computational wavefunctions $\ket{\Psi_\xi}$ includes both the emitters and quantum photons, called ''virtual'', in contrast to the ''real'' photons corresponding to the $\bm{\alpha}_\xi$. To arrive at the partial density matrix of the emitters, we average the conditional projectors, themselves projected onto a quantum vacuum of the virtual photons. That is to say, the reason the virtual photons are called virtual is that their presence is not directly observable by a distant classical detector.

\begin{equation}
    \hat{\rho}_a := \frac{1}{N_\text{batch}} \sum_\xi \frac{\bra{\boldsymbol{0}}\cdot\ketbra{\Psi_\xi}\cdot \ket{\boldsymbol{0}}}{\left|\braket{\Psi_\xi}{\boldsymbol{0}}\right|^2}
\end{equation}

This density matrix can now be used to calculate the expectation value of every emitter-side operator $\hat{O}$:

\begin{equation}
\expval{O}_\text{stoch} := \text{tr}\left(\hat{O} \hat{\rho}_a\right) \approx \expval{\hat{O}}
\end{equation}

The observables of the photonic state are defined by the classical statistics of $\bm{\alpha}$. As such, for an arbitrary antinormally-ordered quantum average characterized by orders $m_1, m_2, ...$, $n_1, n_2, ...$ for modes $\mu_1, \mu_2, ...$, $\nu_1, \nu_2, ...$, there's a stochastic average which converges to the correct quantum value:
\begin{equation}\label{stoch-avg}
 \frac{1}{N_\text{batch}} \sum_\xi \left(\prod_l \alpha_{\xi; \mu_l}^{m_l} \prod_l (\alpha_{\xi; \nu_l}^{n_l})^*\right) \approx \expval{\prod_l \hat{a}_{\mu_l}^{m_l} \prod_l (\hat{a}_{\nu_l}^{n_l})^\dagger}
 \end{equation}
This fact allows us to recover photonic observable averages. Of particular interest to us are the $g^{(n)}$ factors defined as: 

\begin{equation}
    G^{(n)} := \expval{\left(\hat{b}^\dagger\right)^n \hat{b}^n}
\end{equation}

\begin{equation}
    g^{(n)} := G^{(n)}/(G^{(1)})^n
\end{equation}

Stochastically, these are calculated as averaged functions of the real field value $f_\xi$

\begin{equation}\label{eq:f-xi}
    f_\xi := \sum_{\nu=1}^{N_m} c_\nu \alpha_{\xi; \nu} \sqrt{\Delta \omega_\nu}
\end{equation}

\begin{equation}
    G^{(n)} = \frac{1}{N_\text{batch}} \sum_\xi \sum_{m=0}^n |f_\xi|^{2m} (-1)^{n-m} \binom{n}{m} P^n_m
\end{equation}

In addition, we also calculate the average virtual field value registered by the emitters. As such, we also use the emitter-photon density matrix:

\begin{equation}
    \hat{\rho}_{ap} := \frac{1}{N_\text{batch}} \sum_\xi \frac{\ketbra{\Psi_\xi}}{\braket{\Psi_\xi}}
\end{equation}

We then calculate the virtual photon field expectation value as:

\begin{equation}
    F_\text{virt} = \text{Re }\text{tr}\left(\hat{\rho}_{ap} \hat{b}\right)
\end{equation}

Finally, we use Pearson correlations in Fig.~\ref{fig:corr}. The Pearson correlation between two variables is defined in the conventional way:

\begin{equation}\label{eq:def-pears}
    C(x, y) = \frac{\expval{(x - \expval{x})^* (y - \expval{y})}}{\sqrt{\expval{|(x - \expval{x})|^2}}\sqrt{\expval{|(y - \expval{y})|^2}}}
\end{equation}

The quantity plotted is then the absolute value $|C(f(t), f(t'))|$ for the different field values $f$ obtained at different time points $t, t'$ as given by (\ref{eq:f-xi}). We calculate the stochastic averages in (\ref{eq:def-pears}) with respect to the stochastic trajectory index $\xi$.

\section{Numerical implementation}

The resulting TDSE is a system of coupled nonlinear ordinary differential equations which amounts to a slight rewriting of the ones given in \cite{polyakov2019dressed} for purposes of easier computational treatment. The basic Hamiltonian coincides with the one given in (\ref{eq:hgen}). However, it is supplemented by an additional term describing interaction with a stochastically-sampled classical force $\bm{\alpha}$. The advantage of this scheme over exact diagonalization is that in many cases~\cite{hartmann2017exact, polyakov2019dressed}, this additional classical interaction reduces the number of quantum photons involved in the resultant evolution, shrinking drastically the dimension of their Hilbert space.
\begin{equation}
 \hat{H}_0 (t) = \hat{H}_S (t) + \hat{H}_I + \hat{H}_B 
\end{equation}
\begin{equation}\label{ham-stoch}
 \hat{H} [\bm{\alpha}, \ket{\psi}] (t) = \hat{H}_0 (t) + \sum_{\nu = 1}^{N_m} c_\nu^* \alpha_\nu^* \hat{S}_- - \expval{\hat{S}_-}^*_\psi \int dk c(k) \hat{a}_k 
\end{equation}

\begin{equation}\label{syseq}
    \begin{cases}
        i \partial_t \ket{\psi_\xi (t)} &= \hat{H}[\bm{\alpha}_\xi (t), \ket{\psi_\xi (t)}] (t) \ket{\psi_\xi (t)}\\
  i \partial_t \alpha_{\nu; \xi}(t) &= \omega_\nu \alpha_{\nu; \xi}(t) + c_\nu^* \expval{\hat{S}_-}_{\psi_\xi}
    \end{cases}
\end{equation}

The absence of a $+\text{h.c.}$ in (\ref{ham-stoch}) is not in error. As opposed to describing the wavefunction of the full system $\ket{\Psi}$ like TDSE normally does, (\ref{syseq}) deals with a conditional wavefunction for the reduced system $\ket{\psi_\xi(t)}$, measured in coincidence with its coherent state $\bm\alpha_\xi$. As such, its evolution is non-unitary -- and the resulting Hamiltonian non-Hermitian. Consequently, the resulting stochastic wavefunctions $\ket{\psi_\xi}$ are also not normalized. Expectation values in the form $\expval{O}_\psi$ must be understood as normalized averages over vacuum-projected computational wavefunctions:
\begin{equation}
 \expval{\hat O}_\psi \equiv \frac{\braket{\psi}{\bm{0}} \hat{O} \braket{\bm{0}}{\psi}}{|\braket{\psi}{\bm{0}}|^2} 
\end{equation}
(\ref{syseq}) is a system of nonlinear ordinary differential equations for $\ket{\psi}$ and $\bm{\alpha}$ which can still be linearised in a straightforward way. The commonly used approach to solving problems of this kind are the so-called exponential Rosenbrock-type methods \cite{caliari2009implementation, hochbruck2010exponential}. 

We solve the respective TDSE specified by starting at the initial conditions:

\begin{equation}
    \begin{cases}
        \ket{\psi_\xi (-\infty}) &= \ket{g}\\
        \bm{\alpha}_\xi (-\infty) &\sim \mathcal{C N}(\bm{0}, \bm{1})
    \end{cases}
\end{equation}

For streamlining purposes, the solutions for the different initial conditions are stored as multidimensional tensors and solved jointly. The solution-dependent coefficients proportional to e.g. $\alpha_{\nu; \xi}$ and $\expval{\hat{S}_-}_{\psi_\xi}$ are applied after matrix multiplication. 

The selected solver algorithm is the Rosenbrock-Euler scheme. The $\varphi$-functions required for the Rosenbrock method's operation are computed as Taylor expansions up to the floating point error. All calculations are done in FP32 precision. The fixed time step is set to $\Delta t = 0.01$. For these parameters, a photonic band with $N_p=6$ virtual photons and $N_m = 12$ discretized modes, as well as $N_e=40$ atoms as per the main text, the runtime of a solution encompassing 256 trajectories we attain on a single NVIDIA A100 GPU is approximately 265 minutes.

\section{\label{app:sde}Markovian decay channels}

The scheme we apply for simulating non-Markovian dynamics of light emission already involves propagating the stochastic TDSE for large batches of wavefunctions driven by varying Hamiltonians. It would be tempting to reduce the Markovian dynamics to stochastic averages, meaning that instead of having to evolve large batches of density matrices, we would be able to instead keep working with large batches of wavefunctions -- even if we would need more for convergence. And indeed, a density matrix evolving according to the quantum master equation can be represented as the covariance matrix $\mathrm{E}\left[\ketbra{\psi}\right]$ of a stochastic process defined for a wavefunction $\ket{\psi}$. However, the resulting equation, as given in \cite{breuer2002theory}, eq. (6.25), is piecewise deterministic, meaning that it involves discontinuous jumps sampled as increments of the discrete Poisson process $dN$:

\begin{align}\label{eq:pdp}
    d\ket{\psi} = &-i dt \left(\hat{H} + \frac{i \gamma}{2} \left\|\hat{S}_- \ket{\psi}\right\|^2\right) \ket{\psi}\\ &+ \left(\frac{\hat{S}_- \ket{\psi}}{\left\|\hat{S}_- \ket{\psi}\right\|^2} - \ket{\psi}\right) d N \nonumber
\end{align}

Given how these jumps occur at different times for the different wavefunctions of the batch, implementing a batched piecewise deterministic equation naïvely makes batched propagation of TDSE infeasible. However, in some cases, the strict equation (\ref{eq:pdp}) allows for an approximate form involving the Wiener, instead of the Poisson, process. As explained below, existing numerical methods can be leveraged for stochastic differential equations with terms proportional to Wiener increments $d W$. The approximation involved is called the diffusion limit. The resulting equation is (6.214) from \cite{breuer2002theory}:

\begin{align}\label{eq:lindstoch}
    &d \ket{\psi} = -i dt \hat{H} \ket{\psi} \nonumber\\ &- i \gamma dt \left(\frac{1}{2} \hat{S}^\dagger_- \hat{S}_- + \frac{1}{2} \left|\expval{\hat{S}_-}\right|^2 - \expval{\hat{S}_-}^* \hat{S}_-\right)\ket{\psi} \\ &+ \sqrt{\gamma} d W \left(\hat{S}_- - \expval{\hat{S}_-}\right) \ket{\psi}\nonumber
\end{align}

The equation~(\ref{eq:lindstoch}) can be simulated by using a modification of the Euler-Rosenbrock solver. Due to not admitting a finite time derivative, the Wiener process does not allow a treatment such as the Euler-Rosenbrock solver. However, one can derive a correction to the Rosenbrock-Euler method. The method we choose is the Stochastic Exponential Rosenbrock Scheme (SERS), as formulated in~\cite{mukam2018strong}. 

By employing this method, we simulate an additional, smaller dataset consisting of 8192 trajectories, choosing $\gamma/\omega_0 = 10^{-3}$. For a fixed time step $\Delta t = 0.01$, the average time to compute a batch of 256 trajectories is 495 minutes.

\section{\label{app:gn}Photon count statistics}

\begin{figure}[H]
    \centering
    \includegraphics[width=\linewidth]{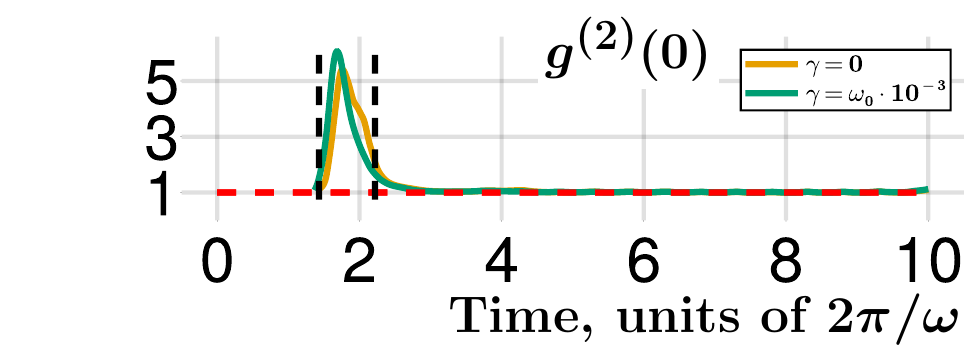}
    \caption{(color online) Photonic $g^{(2)}$ factor $g^{(2)} = \expval{\left(\hat{b}^\dagger\right)^2 \hat{b}^2}/\expval{\hat{b}^\dagger \hat{b}}^2$ plotted with respect to time, measured in laser cycles. After a brief spike typical for a superradiant burst, $g^{(2)}$ falls back to approximately the classical value of 1;}
    \label{fig:gn}
\end{figure}

\begin{figure}[h!]
    \centering
    \includegraphics[width=\linewidth]{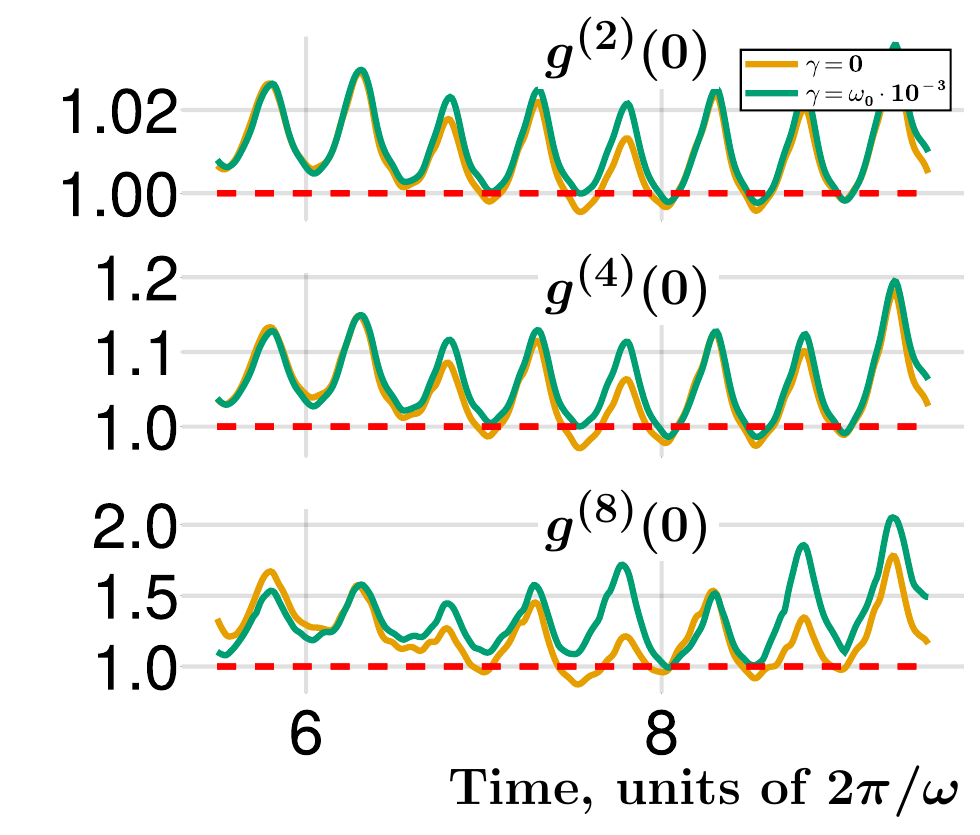}
    \caption{(color online) Photonic $g$ factors $g^{(n)} = \expval{\left(\hat{b}^\dagger\right)^n \hat{b}^n}/\expval{\hat{b}^\dagger \hat{b}}^n$ plotted with respect to time, measured in laser cycles. Here, the time scale is shortened to demonstrate the finer features of temporal evolution of the quantum field in Phase III, shown on a coarse time scale in Fig.~\ref{fig:multiplot-2h}.}
    \label{fig:gn-sup}
\end{figure}

\begin{figure}[H]
    \centering
    \includegraphics[width=\linewidth]{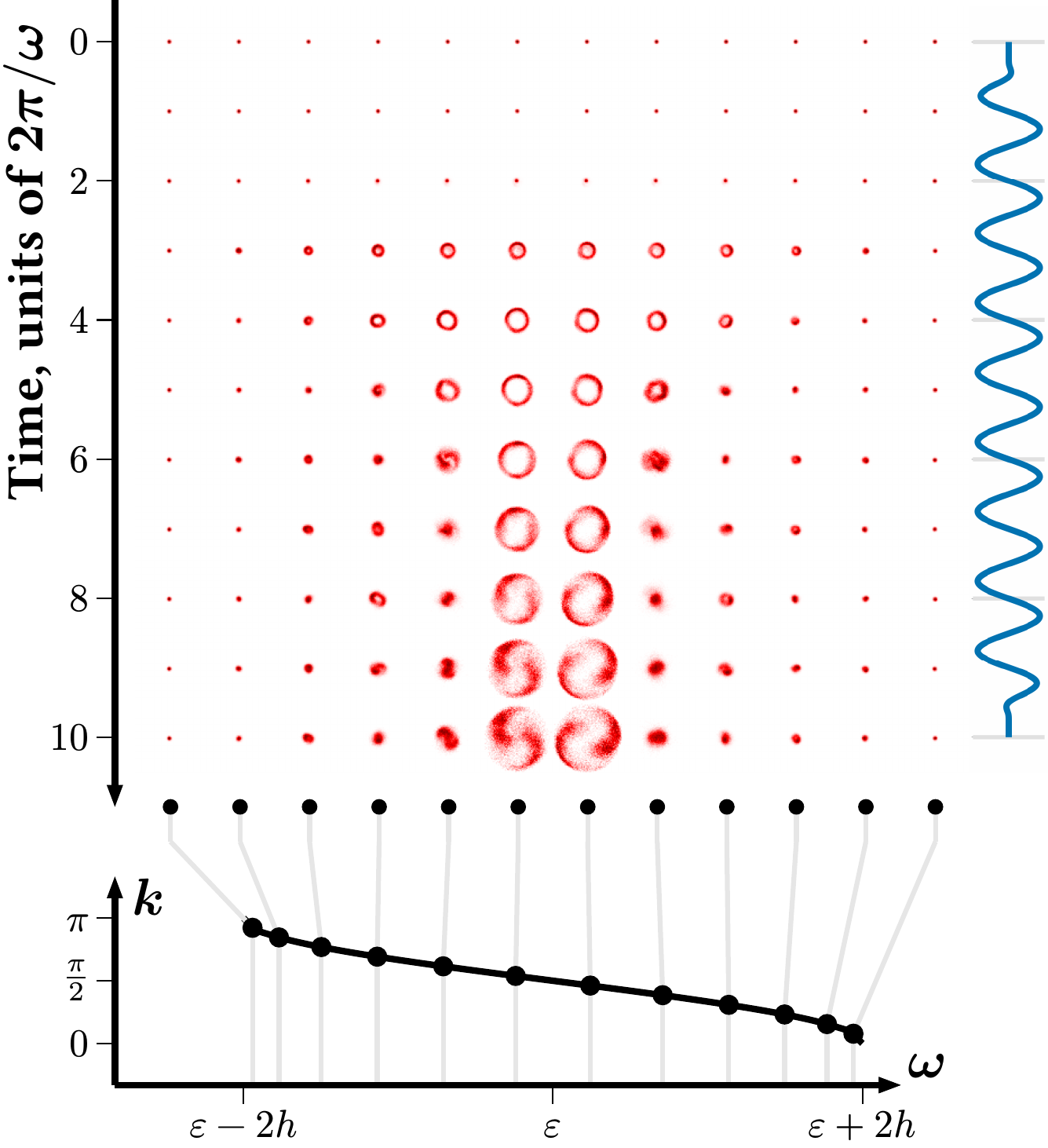}
    \caption{(top) Development of the multimode Husimi function for the different frequency modes left to right over the simulation time (top to bottom, in units of 1 laser cycle $2\pi/\omega$). While some modes display a ring-like Husimi function even on their own, in isolation their photon number dispersion remains decidedly super-Poissonian. (bottom) Discretized modes whose Husimi functions are displayed on top, the modes shown by circles and the exact dispersion relation (\ref{omega-k}) by the solid black line. }
    \label{fig:husgrid-sup}
\end{figure}

The Figures~\ref{fig:gn}, \ref{fig:gn-sup} and \ref{fig:husgrid-sup} provide a general overview of the time-resolved statistics of the radiation emitted by our system. In Fig.~\ref{fig:gn} we observe a typical superradiant setup, with the $g^{(2)}$ factor rising briefly, then dropping back down to the classical value of 1.

However, Fig.~\ref{fig:husgrid-sup} shows that the actual emission statistics are vastly different from conventional superradiance. Instead of a coherent state expected for a continuously-pumped ensemble of atoms, here we observe ring-like Husimi functions form, reminiscent of a Fock state. Indeed, analysing their statistics around the equilibrium shown in Fig.~\ref{fig:gn-sup} reveals mild non-classicality in the photon count statistics. This non-classicality is small due to the scale of the state in question, involving hundreds of photons emitted, as well as the residual entanglement persisting in the equilibrium as seen in Fig.~\ref{fig:multiplot-2h}(d). However, it is numerically significant, i.e. the simulations are converged with respect to the number of discretized modes $N_m$, virtual photons $N_p$, the number of samples $N_\text{batch}$, the error tolerance of the differential equation solver, and the choice of single vs. double floating-point precision. Likewise, it persists in the presence of limited Markovian decoherence, as shown by the green lines in Figures~\ref{fig:gn} and \ref{fig:gn-sup}.

\end{document}